\documentclass[aps,prb,twocolumn,showpacs,citeautoscript,floatfix]{revtex4-1}

\usepackage{graphicx}
\usepackage{amssymb}
\usepackage{amsmath}

\newcommand{\ud}{\,{\mathrm d}}
\newcommand{\zc}{\overline{z}}
\newcommand{\wc}{\overline{w}}
\newcommand{\fc}{\overline{f}}

\def\be{\begin{equation}}
\def\ee{\end{equation}}

\begin{document}

\title{Magnetic states in multiply-connected flat nano-elements.}

\author{Andrei B. Bogatyr\"ev}
\affiliation{Institute for Numerical Mathematics, Russian Academy of Sciences, 8 Gubkina str., Moscow GSP-1, Russia 119991}

\author{Konstantin L. Metlov}
\affiliation{Donetsk Institute of Physics and Technology, 72 R. Luxembourg str., Donetsk}
\email{metlov@fti.dn.ua}
\date{\today}

\begin{abstract}
Flat magnetic nano-elements are an essential component of current and
future spintronic devices. By shaping an element it is possible to
select and stabilize chosen metastable magnetic states, control its
magnetization dynamics. Here, using a recent significant development in
mathematics of conformal mapping, complex variable based approach to the 
description of magnetic states in
planar nano-elements is extended to the case when elements
are multiply-connected (that is, contain holes or magnetic anti-dots).
We show that presence of holes implies a
certain restriction on the set of magnetic states of nano-element.
\end{abstract}
\maketitle

While existence of topological solitons\cite{S58} as metastable states
in infinite 2-d ferromagnets was known theoretically for quite a long
time\cite{BP75,Woo77,G78}, real systems have finite size. Their boundary
imposes restrictions on the set of low-energy topological
states, making it equivalent to the set of rational functions
of complex variable with {\em real} coefficients\cite{M10}, as opposed 
to {\em complex} coefficients in the laterally unconstrained thin film case\cite{BP75}. 
Also, restricted geometry implies the possibility of
formation of half-vortex states, pinned at the side of the planar 
magnet\cite{M01_solitons2,M10}, which are topologically similar to the boundary 
states in the fractional quantum Hall systems and topological insulators.
The shape of nano-element enters this complex description\cite{M10}
via its conformal mapping to the unit disk. 
It is also possible to express Landau-Lifshitz magnetization
dynamics directly in terms of these complex functions using Lagrangian 
approach\cite{M13.dynamics}. The complex description of 
magnetization states\cite{M10} and their dynamics\cite{M13.dynamics,M13.frequency} 
constitutes a complete set of analytical tools to study magnetic textures in simply-connected
nanomagnets far from magnetic saturation. Here this description is extended for the case 
when planar nano-magnet is {\em multiply-connected}.

Let us briefly introduce the description for simply-connected case\cite{M10}. In
small enough magnets the surface effects dominate the volume ones, and
also the exchange interaction is more important than the magnetostatic
one. The former is typical for any small systems, while the latter is 
easy to understand based on the representation of the magnetostatic
energy in terms of the interaction of fictitious magnetic charges of 
opposite signs. The  total magnetic charge is always zero, thus, whatever the 
distribution of the magnetization is, as the size of the magnet 
decreases, the positive and negative charges are brought closer together, 
so that their positive self-energy gets more and
more compensated by their negative mutual interaction energy. The scaling of the exchange energy 
of the set of such an arbitrary magnetization distributions follows the 
volume of the magnet and does not have such an additional reduction. Thus,
in small magnets the exchange interaction becomes more important. 
To make use of this energy hierarchy, we build an approximate expressions
for magnetic states by minimizing the energy terms sequentially (as opposed to 
minimization of their sum, which would result in the exact theory). The process
of sequential minimization is analogous to sieving.

First, consider
all possible magnetization distributions with fixed length of
magnetization vector $|\vec{m}|=1$:
\begin{eqnarray}
m_X+\imath m_Y = \frac{2 w(z,\zc)}{1+w(z,\zc)\wc(z,zc)} \label{eq:magveccomp} \\
m_Z = \frac{1-w(z,\zc)\wc(z,zc)}{1-w(z,\zc)\wc(z,zc)}, \nonumber
\end{eqnarray}
where $\vec{m}=\{m_X,m_Y,m_Z\}=\vec{M}/M_S$ is the local magnetization
vector, expressed in units of saturation magnetization $M_S$;
$z=X+\imath Y$, with $X$, $Y$ and $Z$ being the Cartesian coordinates
(the element is assumed to be the a flat cylinder with axis, parallel
to $Z$ and magnetization distributions are assumed to be
$Z$-independent), $w(z,\zc)$ is a function (not necessarily
meromorphic) of complex variable $z$.

Second, out of all these functions select the ones, which, additionally to having the constant
length of the magnetization, 
minimize the exchange energy (these are the famous Belavin and Polyakov solitons\cite{BP75}). 
Then further restrict the set of the remaining functions twice
by selecting those, which give minimum for the energies of face magnetic charges and those, 
which minimize (in fact, totally avoid) the side magnetic charges.
The final result\cite{M10} of such a selection will be the following representation for
$w(z,\zc)$
\begin{equation}
  \label{eq:solSM}
  w(z,\overline{z})=\left\{
    \begin{array}{ll}
      f(z)/e_1 & |f(z)| \leq e_1 \\
      f(z)/\sqrt{f(z) \fc(\zc)} & e_1<|f(z)| \leq e_2\\
      f(z)/e_2 & |f(z)| > e_2
    \end{array}
    \right. ,
\end{equation}
where $e_1$ and $e_2$ are real positive constants. The complex 
function $f(z)$ is a solution of Riemann-Hilbert boundary value
problem of finding the meromorphic function in the domain ${\cal D}$, 
which corresponds to the face of the planar nanoelement,
having no normal components to the domain's boundary.
This problem usually has many solutions. For simply-connected 
case their set is equivalent to the set of rational functions with real coefficients (constants), 
whose zeros(poles) correspond to vortex(anti-vortex) centers\cite{M10}. 

For example, in a unit disk the subset of states with no antivortices
can be expressed\cite{M01_solitons2} as
\begin{equation}
 \label{eq:simplydisk}
 f^\mathrm{disk}(z)=\imath z c + A - \overline{A} z^2,
\end{equation}
where $c$ and $A$ are a real and a complex constants respectively. While the equilibrium values 
of these constants can be found only by minimizing the total (including the 
energy of the volume magnetic charges) energy
of the nanomagnet\cite{ML08}, the expression (\ref{eq:simplydisk}) can already be useful to pinpoint that 
there are several types of magnetic states in the disk, such as centered 
magnetic vortex (when $A=0$), the so-called ``leaf'' state (when $c=0$) and 
$C$-like magnetization state (when $4 A\overline{A}>c$). These states are shown in Figure 1.
\begin{figure}
\includegraphics[scale=0.5]{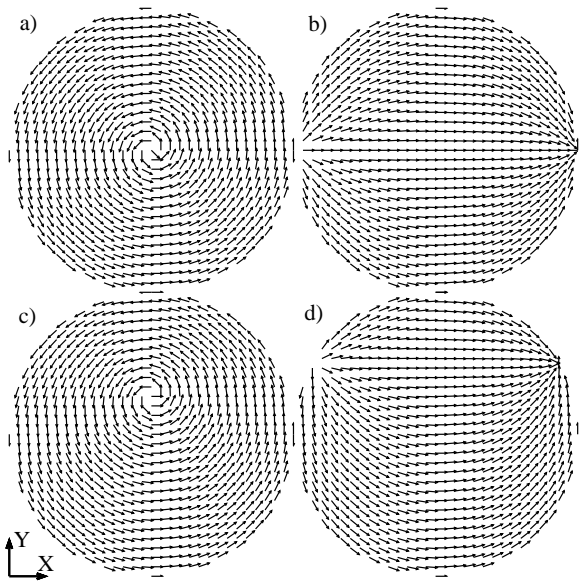}
\caption{\label{fig:disk} Equilibrium and transient magnetization states in ferromagnetic 
nano-disk following from the Eq.~\ref{eq:simplydisk}: a) centered magnetic vortex;
b) ``leaf'' state; c) displaced magnetic vortex; d) ``C''-like state.}
\end{figure}
The solution (\ref{eq:simplydisk}) can be extended when there are additional vortex-antivortex 
pairs present\cite{M10}, allowing to describe both transient dynamical states as well as finite
(simple and topologically charged domain walls).

Before proceeding to the consideration of multiply-connected case, let us
note a general property of the above Riemann-Hilbert boundary
value problem. If it has a solution $f(z)$ in the region $z\in{\cal D}$, the 
corresponding solution $u(t)$ in another region $t\in{\cal D'}$, 
connected to ${\cal D}$ via the conformal 
mapping $t=T(z)$, can be expressed as
\begin{equation}
 \label{eq:conformal}
 u(t) = \left. f(z) T'(z) \right|_{z=T^{(-1)}(t)}, 
\end{equation}
where $T^{(-1)}(t)$ is the inverse of the conformal mapping $t=T(z)$,
which is always defined. This expression applies both to simply- 
and multiply-connected cases and
allows to express $u(t)$ in an arbitrarily shaped cylinder, based
on its expressions $f(z)$ for some canonical domains.
In the simply-connected case these canonical domains can be chosen, based on
convenience only, to be the unit disk\cite{M01_solitons2}, the 
half-plane\cite{M10} or any other simply connected region for which
the solution can be written explicitly. Selection of the canonical domain 
is trickier in the multiply-connected case, since not
every pair of regions with the same connectivity can be conformally
mapped into each other.

\begin{figure}
\includegraphics[scale=0.4]{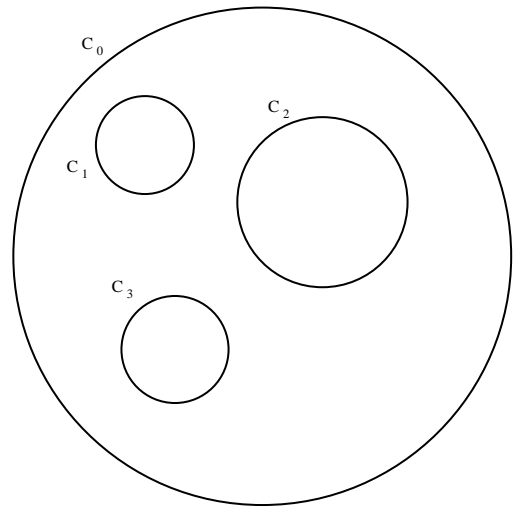}
\caption{\label{fig:circulardomain}An example of a quadruply-connected circular domain with an outer circular 
boundary $C_0$ and inner boundaries $C_j$, $j=1,2,3$.}
\end{figure}

Thanks to the Koebe's theorems, it is well known that
it is possible to define parametrized canonical families of multiply-connected
regions, which, after parameter adjustment, can be mapped to an arbitrary region of
the same connectivity. Among them is the family\cite{Koebe1914} of circular domains with 
cut out inner circles shown in Fig.~\ref{fig:circulardomain}.
Positions and radii of the circular holes in these domains 
are not arbitrary and are dictated by the shape of the target multiply-connected region ${\cal D'}$ to 
which $f(z)$ may be mapped. Because such canonical regions by themselves have physical
significance (such as the simplest case of concentric ring, which is doubly-connected), 
let us choose them as the basis and set the goal of expressing the solutions $f(z)$ of 
the Riemann-Hilbert problem in these regions.

But first we need to address the question of
existence of such solutions and their general properties. Formally, given a 
finitely connected flat domain ${\cal D}$ with good enough boundary (say,
piecewise analytic), the problem
is to find a meromorphic function $f(z)$ in the domain, satisfying the
boundary condition 
\be
\label{RH}
Re(f(z)\overline{n(z)})=0,
\qquad z\in\partial{\cal D},  
\ee
where $n(z)$ is a normal to the boundary of the domain. The solution $f(z)$ may have
zeros and polynomial singularities at the boundary. This (Riemann-Hilbert) boundary 
condition can be reformulated as follows:
\be
\label{RealDiff}
Im~ \frac{\ud z}{f(z)} =0,
\qquad z\in\partial{\cal D},  
\ee
which means that meromorphic differential $(\ud z)/f(z)$ is \emph{real} (i.e. the
restriction of this differential to the boundary is a real differential). The
latter condition is conformally invariant, which, in particular,
implies (\ref{eq:conformal}).

To describe the set of all real differentials in ${\cal D}$, the useful notion
is the \emph{double} of the domain.
The latter is a compact Riemann surface $X({\cal D})$ of genus $g$ equal to the
number of the boundary components of ${\cal D}$
minus one. It is made of two copies of the domain glued along its boundaries,
$z$ being the conformal coordinate on one copy and 
$\bar{z}$ -- on the other. This surface admits natural \emph{anticonformal
involution} $\bar{J}$ (or \emph{reflection}), the interchange of copies, which
is stationary exactly on the boundary components of the domain ${\cal D}$. The
reflection acts on the differentials and the condition for the differential
$d\eta$ to be real is exactly the following:
$$
\bar{J}\ud\eta=\overline{\ud\eta}.
$$ 

{\bf Corollary:}
The following topological phenomenon can be observed:
$$
\sharp\{f(z)~poles\}-\sharp\{f(z)~zeroes\}=\sharp\{{\cal D}~boundaries\}-2
$$
where the poles and the zeros are counted with their multiplicities and for those
at the boundary the multiplicity should be divided by two.

{\bf Proof:}  The solution $f(z)$  corresponds to the meromorphic differential
$\ud\eta:=(\ud z)/f(z)$
on the double of ${\cal D}$. The degree of the divisor of a differential is
$2g-2$.

This implies that we can't control independently number of
zeros and poles of $f(z)$. 

The natural recipe to cook a real differential is symmetrization: take any
meromorphic differential $\ud\eta$ on the surface,
then its symmetrization $\ud\eta+\bar{J}\overline{\ud\eta}$ is meromorphic and real.
This recipe is easy to use once we have a 
representation of the double $X({\cal D})$ as an algebraic curve. This representation, 
for the case of circular multiply connected domains can be represented via a series over
the elements of the corresponding Schottky group\cite{S1877}. The evaluation of such series, however, can
be very inconvenient if approached directly. That's why in the following we will give the 
representation of the solution  $f(z)$ via the Schottky-Klein prime function\cite{Schottky1887,Klein1890}, 
which not only admits an efficient numerical evaluation\cite{CrowdyMarshall2007}, but can also be directly
used in building the conformal map $t=T(z)$ of ${\cal D}$ to ${\cal D'}$ if the latter 
is a multiply-connected polygonal domain\cite{Crowdy2005}. It also has a number of applications in 
problems of optimization and computation\cite{Bogatyrev2007,Bogatyrev2009}.

Informally (for the formal definition see e.g. Ref.~\onlinecite{CrowdyMarshall2007}), 
the Schottky-Klein prime function $w(z,\zeta)$ for a specific multiply-connected circular
domain ${\cal D}$ can be thought as a generalization of the difference
\begin{equation}
 z-\zeta=w_1(z,\zeta),
\end{equation}
which in multiply connected case becomes
\begin{equation}
 w=w(z,\zeta).
\end{equation}
In a simply-connected case products of such differences can be used to build the rational functions of complex variable,
such as those equivalent to the set of Belavin-Polyakov solitons\cite{BP75} or a similar set of states of
finite nano-magnet\cite{M10}, or those, entering the
Schwarz-Christoffel formula for the conformal mapping of polygons. Generalization of the latter to the multiply-connected
case was done by Crowdy\cite{Crowdy2005}, here we shall outline a generalization of the former.

In the simplest doubly-connected case of a concentric ring with an outer radius of $1$ and an inner radius of $q<1$ 
the Schottky-Klein prime function can be expressed as a product
\begin{equation}
 w_r(z,\zeta)=(z-\zeta)\frac{\prod_{k=1}^\infty (1-q^{2 k} z/\zeta)(1-q^{2 k}\zeta/z)}
 {\left(\prod_{k=1}^\infty (1-q^{2 k})\right)^2}
\end{equation}
or written via the q-Pochhammer symbols $(a;q)_n$ as
\begin{equation}
 w_r(z,\zeta)=\frac{z\zeta}{z-\zeta} 
 \frac{(\zeta/z,q^2)_\infty\,(z/\zeta,q^2)_\infty}
      {\left((q^2,q^2)_\infty\right)^2}.
\end{equation}
This representation will be used in some of the following examples.

To build the solutions of the Riemann-Hilbert problem with the specified positions of vortices (or 
anti-vortices) let us define two auxiliary functions,
following from those, introduced in the Section 5 of Ref.~\onlinecite{Crowdy2005}:
\begin{eqnarray}
 F_1(z,\zeta_1,\zeta_2)&=&\frac{w(z,\zeta_1)}{w(z,\zeta_2)}, \\
 F_2(z,\zeta)&=&\frac{w(z,\zeta)w(z,1/\overline{\zeta})}{w(z,\overline{\zeta})w(z,1/\zeta)},
\end{eqnarray}
which,
\begin{figure}[htbp]
\includegraphics[scale=0.5]{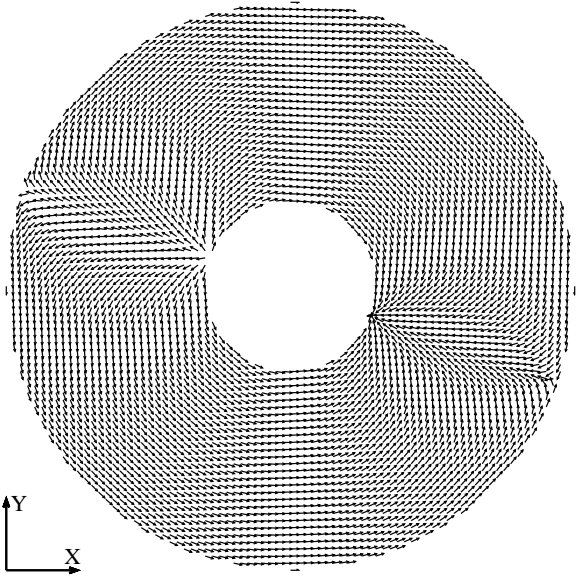}
\caption{\label{fig:finitewall}Finite domain walls in a ring of $q=0.3$, given by (\ref{eq:magveccomp}) and (\ref{eq:solSM}) 
with $e_1=0$, $e_2\rightarrow\infty$, $g(z)=(\partial/\partial z) \log F_1(z,e^{-8\pi \imath/9},e^{\pi \imath/9})$ and
$f(z)=g(1/z)$.} 
\end{figure}
\begin{figure}[htbp]
\includegraphics[scale=0.5]{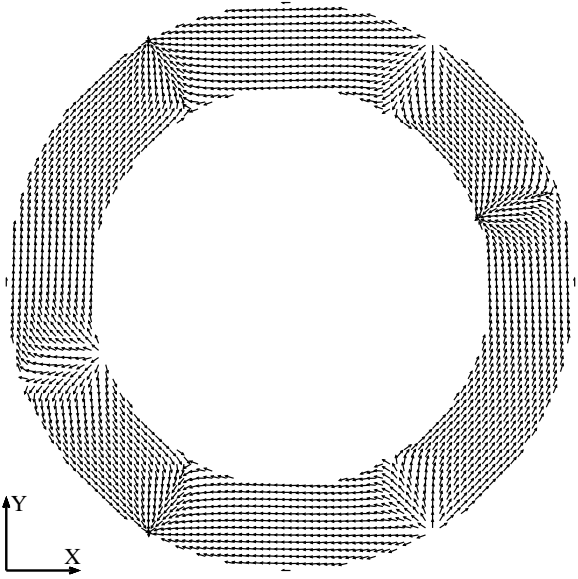}
\caption{\label{fig:domains}A domain structure with several finite domain walls in a ring of $q=0.7$, with $g(z)$ 
containing now the product of three $F_1$ functions: $F_1(z,e^{8\pi \imath/9},e^{-\pi \imath/9})$,
$F_1(z,q e^{4\pi \imath/3},q e^{-2\pi \imath/3})$ and $F_1(z,q e^{\pi \imath/3},q e^{-\pi \imath/3})$. The rest of
parameters are the same as in Figure~\ref{fig:finitewall}.}
\end{figure}
\begin{figure}[htbp]
\includegraphics[scale=0.5]{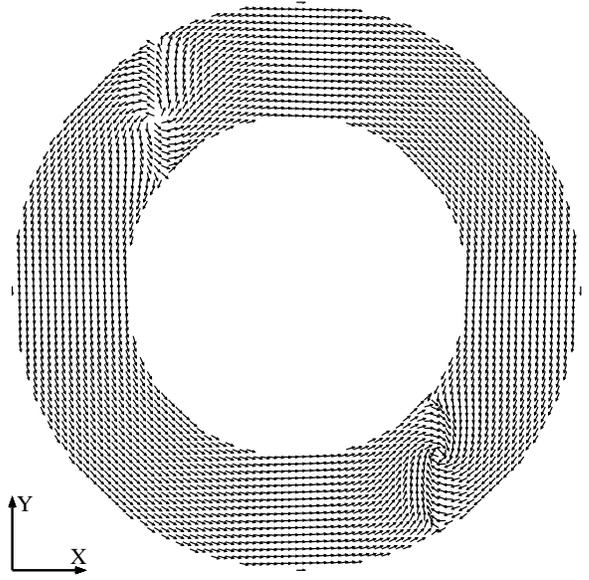}
\caption{\label{fig:vortexwall}Vortex domain walls in a ring of $q=0.6$, with $g(z)$ containing the
product of $F_1(z,e^{4\pi \imath/3},e^{\pi \imath/3})$ and $F_1(z,q e^{11 \pi \imath/9},q e^{2 \pi \imath/9})$. 
The rest of parameters are the same as in Figure~\ref{fig:finitewall}.}
\end{figure}
\begin{figure}[htb]
\includegraphics[scale=0.5]{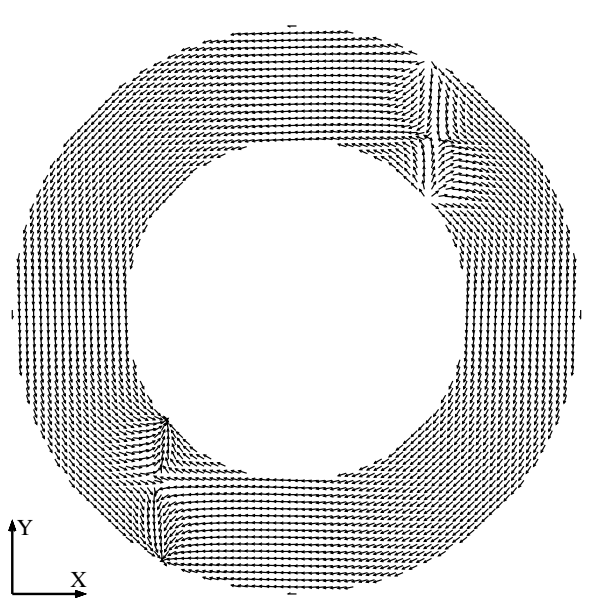}
\caption{\label{fig:vortexantiwall}Antivortex domain walls in a ring. The parameters for this plot are the same
as in Figure~\ref{fig:vortexwall}, except that now $f(z)=g(z)$.}
\end{figure}
\begin{figure}[htb]
\includegraphics[scale=0.5]{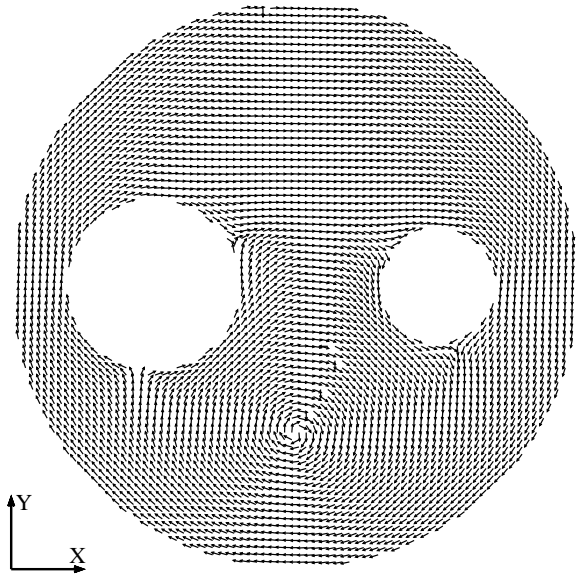}
\caption{\label{fig:triplyconnected}A vortex in a triply connected domain, with the cut out circles of the radii 
0.2 and 0.3, located at x=0.5 and x=-0.5 respectively. The function 
$g(z)=(\partial/\partial z) \log F_2(z,-0.52\imath)$ and $f(z)=g(z)$. The Schottky-Klein prime function was 
evaluated numerically, using the Matlab package developed in Ref.~\onlinecite{CrowdyMarshall2007}.}
\end{figure}
provided $\zeta_1$ and $\zeta_2$ belong to the same inner or outer circle  $C_j$ (with $j=0,1,2,\ldots$) 
of the multiply-connected circular domain and $\zeta$ is any point inside of it, have constant argument 
(complex phase) 
on all the boundaries $C_j$ of the said domain. This means that any product of these functions will have 
the constant argument on the boundaries of the circular domain too, which implies that the logarithmic 
derivative of this product
\begin{equation}
 g(z)=\frac{\partial}{\partial z} \log \left(\prod_m F_1(z, \zeta_{1,m}, \zeta_{2,m}) \prod_n F_2(z, \zeta_n)) \right)
\end{equation}
will be an analytic function of $z$ satisfying the condition (\ref{RH}) at the boundary of the multiply-connected
circular domain. That is, $f(z)=g(z)$ will be a solution of the Riemann-Hilbert problem with the specified (by the parameters 
$\zeta_{1,m}$, $\zeta_{2,m}$, $\zeta_n$) positions of zeros and poles. It will have some additional zeros and poles too,
so that the previously mentioned topological constraint is always satisfied. Also, it is easy to show that the function 
$f(z)=g(1/z)$ corresponds to a ``mirror'' solution of the problem, where the vortices, specified by the parameters $\zeta$, 
correspond to the antivortices and vice versa. This is illustrated by several examples in 
Figures~\ref{fig:finitewall}-\ref{fig:triplyconnected}, some of which closely resemble well known magnetization textures, 
observed in ferromagnetic nanorings.

From these, the corresponding magnetization textures in the nano-magnets of the same high connectivity but other
shapes can be derived according to the Eq.~\ref{eq:conformal} using the conformal mapping (e.g. the mapping
to the multiply-connected polygonal domains, which are also expressed in terms of 
Schottky-Klein prime function\cite{Crowdy2005}).

Thus, we have built an approximate analytical representation of the low-energy magnetization states in planar
multiply-connected nanomagnets. It is parametrized via the positions of some of vortices or 
antivortices in the magnet. Not all topological singularities can be freely placed. There are additional 
restrictions on their positions, whose number is equal to the number of boundary components minus one.
These restrictions will be discussed in our forthcoming paper. Moreover, there is a ``soft''
constraint on the number of vortices and antivortices in relation to the connectivity of the system, 
present in or near the equilibrium. It is of the same ``soft'' nature as the topological solitons themselves 
in finite systems. Really, only in the infinite 2-d magnet the topological charge is conserved and there is 
an infinitely high energy barrier\cite{BP75}, separating the states with different topological charge. In finite magnets
this barrier is finite and vortices may enter/exit the particle through the boundary, changing the topological
charge of the magnetization texture inside. Yet, once this barrier is holding 
(and the magnetization vector is pinned to the boundary of the nanomagnet) the vortices inside the planar
nanoelements behave like true topological solitons.

Complete solution of the micromagnetic problems, using these parametrized 
magnetization textures as trial functions, will require computation and minimization of the 
total energy of the magnet (including the exchange, magnetostatic
and other energy terms), or, solution of the equations 
of the magnetization dynamics using the parameters as collective variables\cite{M13.dynamics}. 
In many cases this can be a very complex task, which is probably still easier to approach numerically 
(as it was done in the framework of the Magnetism@home distributed computational project\cite{M12.large}). 
Yet, even without the total energy computation, these analytical 
trial functions can still be useful to understand, classify and interpret the magnetization textures,
obtained in the experiment and simulations. Also, they can lead to beautiful analytical results, 
with the generality well beyond what numerical micromagnetics may offer. Like, for example, the formula
for the magnetic vortex precession frequency\cite{GHKDB06}, based on the displaced vortex 
model (\ref{eq:simplydisk}). The presented ``complex variables'' approach might generate more useful
models, now in multiply-connected case as well.


%

\end{document}